\RequirePackage{fix-cm}
\RequirePackage{setspace,amsmath,amsfonts,amssymb}

\documentclass[11pt, a4paper]{article}

\usepackage{graphicx}

\usepackage{rotating}
\usepackage{subcaption}
\usepackage[usenames,dvipsnames]{xcolor}
\usepackage[round, sort]{natbib}
\usepackage{moreverb}
\usepackage{url}
\definecolor{Red}{rgb}{0.5,0,0}
\definecolor{NavyBlue}{rgb}{0.1,0.1,0.45}
\definecolor{MidnightBlue}{rgb}{0.1,0.1,0.65}

\usepackage[
    hyperfootnotes=true, plainpages=false,%
    pdfpagelabels,%
    hypertexnames=true,%
    naturalnames,%
    pdfproducer={Latex},%
    pdfcreator={pdflatex},%
    bookmarks,bookmarksnumbered,
    colorlinks,%
    linkcolor=MidnightBlue,%
    citecolor=NavyBlue,%
    filecolor=black,%
    urlcolor=Red,%
    breaklinks=true%
    ]{hyperref}

\usepackage{lscape}

\newcommand{\Prob}{\mathsf{P}}

\newcommand{\RNum}[1]{\uppercase\expandafter{\romannumeral #1\relax}}

\usepackage{fullpage}
\usepackage{booktabs}
\usepackage{multirow}
\usepackage{rotating}
\usepackage{pifont}
\usepackage{authblk}
\usepackage{setspace}
\usepackage{threeparttable}
\usepackage{tabularx}
\usepackage{pbox}

\begin{document}

\title{Plus-Minus Player Ratings for Soccer}

\author{Tarak Kharrat\footnote{Salford Business School, University of Salford, Manchester. \texttt{T.Kharrat@salford.ac.uk} }        \and
        Javier L\'opez Pe\~na\footnote{Department of Mathematics, University College London. \texttt{j.lopezpena@ucl.ac.uk} }  \and
        Ian McHale\footnote{Salford Business School, University of Salford, Manchester. \texttt{i.mchale@salford.ac.uk} }              
}

\maketitle

\begin{abstract}
  The paper presents a plus-minus rating for use in association football
  (soccer). We first describe the standard plus-minus methodology as used in
  basketball and ice-hockey and then adapt it for use in soccer. The usual
  goal-differential plus-minus is considered before two variations are
  proposed. For the first variation, we present a methodology to calculate an
  \emph{expected goals} plus-minus rating. The second variation makes use of
  in-play probabilities of match outcome to evaluate an \emph{expected points}
  plus-minus rating.  We use the ratings to examine who are the best players in
  European football, and demonstrate how the players' ratings evolve over
  time. Finally, we shed light on the debate regarding which is the strongest
  league. The model suggests the English Premier League is the strongest, with
  the German Bundesliga a close runner-up.
\end{abstract}

\section{Introduction}

In sport, there is great interest in evaluating and measuring the performance of
players. In team sports, owners, managers and coaches want to identify which
players are key to their team's success, so that recruitment and retention of
players can be properly informed. Unlike other industries, there is much
external interest in the performance of a sports team's \emph{employees} from,
for example, fans and the media wanting to know which players to support, and
which to berate.  As such, one of the main tasks of sports analytics is to
evaluate the performance of players and understand their contribution to the
team's results.

In this paper we present two modifications of the well-known \emph{plus-minus}
(PM) ratings model previously used to identify key players in basketball (see, for example,
\citet*{sill2010improved}) and ice-hockey (see, for example, \citet*{macdonald2012adjusted}).
The PM ratings system is simple and intuitive, and provides an answer to the
question: `how does a team perform with a player, compared to without the
player?'. The modifications we propose are specific to soccer - a game in which
it is notoriously difficult to rate players objectively.

For individual sports like tennis and chess, rating players is perhaps
simpler than for team sports. Paired comparisons models are
well-established and several variations
exist. \citet*{mchale2011bradley} provide a ratings system for tennis for example. 
A perhaps more complex task is to estimate time-varying ratings 
for individuals which update following new information (the latest
results). Elo ratings have been used for over half a century for rating
chess players. Similarly, the \emph{Glicko} 
rating system \citep*{glickman2012example}  provides a more theoretically
justified model for estimating time-varying ratings of individuals.

More recently, attention has moved to using machine learning techniques to
estimate player ratings.
The \emph{TrueSkill} rating \citep*{herbrich2007} developed at 
Microsoft is a generalisation of
the Elo ratings and is used for rating video game
players. 

Rating players in sports teams is more problematic.  Players often have
different responsibilities with some concentrated on offence (i.e.~aiding
scoring), whilst others are specialised in defence (i.e.~helping to prevent
scores for the opposition).  A commonly used approach is to assign a value to a
set of actions considered to be `of interest' and to reward the player taking
them with the associated value. This method was used for example in the EA SPORTS Player 
Performance Indicator \citep*{mchale2012development} and is still used by the English Premier League
as the official player ratings system.  Due to its additivity, the previous
approach provides simple, user-friendly player ratings and rankings. However, a
cost of the simplicity is the lack of context and a deeper understanding of the
situations in which actions were committed. Further, the data requirement is not
trivial.

Models have been used to rate players for specific tasks.  For example,
\citet*{saez2013expected} and \citet*{mchale2014mixed} present methods to
identify the scoring ability of footballers whereas \citet*{lopez2012networks},
\citet*{pena2015can}, \citet*{brooks2016developing} and
\citet*{szczepanski2016beyond} deal with the passing aspect.  But identifying
the overall contribution of a player to a team's success (or lack of it) has
proven difficult in soccer. However, the concept of the PM ratings provides
hope.

The concept of the PM rating is fundamentally different to the rating mechanisms
discussed above. It directly measures the contribution a player has on a team's
success as measured by (the differential) of a target metric (goals for
example).  It does not make use of event data, and is not concerned with the
number of actions a player might have achieved. All that matters is
\textit{``whilst the player was on the pitch, what happened to the target
  metric?''}.

The PM rating has been used extensively in ice-hockey
(\citet*{macdonald2012adjusted}; \citet*{spagnola2013complete}) and basketball
(\citet*{sill2010improved}; \citet*{fearnhead2011estimating}).  Indeed, PM
ratings are now part of the statistics reported by the media (ESPN for
example\footnote{\url{http://www.espn.com/nba/statistics/rpm}}) and professional
leagues (since 1968 the NHL has kept track of each player's PM
rating\footnote{\url{http://www.nhl.com/stats/player}}) in US team
sports. Surprisingly, plus-minus ratings have yet to be adopted in soccer and
are only discussed in some specialised
forums\footnote{\url{http://www.soccermetrics.net/player-performance/adjusted-plus-minus-deep-analysis}}
but, to the best of our knowledge, have never been studied in the academic
literature.

In this paper we propose to fill in the gap and adapt the plus-minus rating for
use in soccer. We present the model currently used in basketball, and reported
by ESPN before suggesting modifications to adapt the methodology for use in
soccer. We then propose two extensions of the methodology using new target
metrics measuring team success: first, we present an expected goals (xG)
plus-minus rating (xGPM); and second, we present an expected Points (xP)
plus-minus rating (xPPM).  For the xGPM ratings we use a model to calculate the
probability of a shot resulting in a goal.  For the xPPM ratings we use an
`in-play' model to estimate the probability of each match outcome (win, draw,
loss) given the current game state at any moment during the game.  Both models
are presented below.

The remainder of the paper is structured as follows: First, we describe the data
used for this research (Section~\ref{sec:data}). In Section~\ref{sec:PlusMinus}
we present the basic plus-minus rating and the \emph{regularized adjusted}
plus-minus rating currently used in basketball. In
Section~\ref{sec:PlusMinusForFootball} we describe two new variations of the
plus-minus ratings: an expected goals rating (xGPM), and an expected points
plus-minus rating (xPPM).  In Section~\ref{sec:Testing} we use the ratings to
look for the top players across European soccer, and see how their ratings
evolve over time, before using the model to examine the relative strengths of
European leagues.  We conclude with some closing remarks in
Section~\ref{sec:conclusion}.

\label{sec:data}
\section{Data}
We collected data from 11 European leagues over the last 8 seasons as
detailed in Table~\ref{tab:EuroLeague}. For every game in our data set, we collect the
match date, the starting line-ups, timings of any goals, and timings and player names of any substitutions and red cards.  
\begin{table}[ht]
  \centering
  \begin{tabular}{llr}
    \toprule
     League                 &         Seasons   &  Games \\
    \midrule
    England Premier League  & 2009/10--2016/17  & 3,040 \\ 
    Germany Bundesliga I    & 2009/10--2016/17  & 2,448 \\ 
    Spain La Liga           & 2009/10--2016/17  & 3,039 \\ 
    Italy Serie A           & 2009/10--2016/17  & 3,037 \\ 
    Germany Bundesliga II   & 2015/16--2016/17  & 612 \\ 
    England Championship    & 2013/14--2016/17  & 2,227 \\ 
    Netherlands Eredivisie  & 2013/14--2016/17  & 1,242 \\ 
    Turkey Super Lig        & 2014/15--2016/17  & 918 \\ 
    Portugal Liga NOS       & 2016/17           & 306 \\ 
    France Ligue 1          & 2009/10--2016/17  & 3,039 \\ 
    Russia Premier League   & 2013/14--2016/17  & 960 \\
    \midrule
    \textbf{Total}          &                   & \textbf{20,868}\\
    \bottomrule
  \end{tabular}
  \caption{Description of data used by league and season.}\label{tab:EuroLeague}
\end{table}

For the expected goals model developed in Section~\ref{sec:PMxG}, additional
information regarding shots is needed. Specifically, the shot time, the
shooter $(x,y)$ coordinates, the type of shot (penalty, free-kick, header or
open play), and the subjective ``\textit{big chance}'' qualifier describing
the shot situation are extracted from Opta F24 feed.  On top of that, goal-keeper 
skills as reported by EA SPORTS FIFA video games are collected. They describe the keeper's
\textbf{diving}, \textbf{ball handling}, \textbf{ball kicking},
\textbf{positioning}, and \textbf{reflexes} skills. Mapping players between the
Opta feed and EA SPORTS is done using the Google research tools to match
players' names and using the date of birth and the player's country of birth for
validation. Players not found by this method are mapped manually.

\section{The Plus-Minus Rating}
\label{sec:PlusMinus}
The plus-minus statistic has been in use since the 1950s in ice-hockey but is
most seen nowadays applied to basketball. Indeed, the complexity of the game
of basketball has led to several developments of the original
concept. In this section we will first describe the basic
plus-minus statistic, 
before presenting modifications that have been introduced in the literature.
In what follows we will define everything in terms of soccer.

\subsection{The Basic Plus-Minus Statistic}
In its simplest form, a player's plus-minus statistic can be used to answer the question:
``what happens when the player is on the pitch, compared to when he is off it?''.
Historically, goals (or points in basket ball) have been the preferred way to measure
``what happened'' and the raw plus-minus score calculates the player's contribution to the
goal difference of his team (per ninety minutes) whilst he is on the pitch. For example,
consider a player who makes two match appearances. In the first match, he plays the first
60 minutes during which the team concedes one goal and fails to score itself. The match
finishes in a 1-0 loss. In the second match, the player comes to the field with 30 minutes
remaining and his team is enjoying being 3-0 ahead. During the 30 minutes of play he is on
the pitch, the score moves to 5-0. The player's plus-minus rating is then $(
(-1/60)+(2/30) ) \times 90 = +4.5$. In other words, when the player was on the pitch the
team scored 4.5 goals per 90 minutes more than the opposition.

The net plus-minus statistic can be used to measure the importance of
a player to his team. This is simply the plus-minus statistic when the player is
on the pitch minus the plus-minus statistic when the player
is not on the pitch. In our example, the plus-minus statistic without
the player is $((0/30) + (3/60))\times90=4.5$, so that the
net plus-minus statistic is 0. It appears then that the
team is equally successful with and without the player.

This is of course a very simplistic picture and several pieces of information are not
taken into account. For example, the effects of strengths of the other players on the
pitch or of the game situation (such as a reduction in the number of players on a team
following a red card, or of any home advantage) have not been accounted for. Further, if
one was to use this simple plus-minus rating to compare players from different teams, the
results would be almost meaningless. Consider two different players: one playing for the
league champions and the other playing for the leagues worst team. Suppose both players
had pure plus-minus ratings of 0. Who is likely the better player? Most sports fans would
say that the player achieving a pure plus-minus of 0 on the league's worst team probably
deserves more credit in this example.

To account for these
factors, the adjusted plus-minus statistic was introduced, and is described next.

\subsection{Regularized Adjusted Plus-Minus}
\label{ssec:RegAdjustedPM}
The adjusted plus-minus player metric was first described in
\citet{rosenbaum2004measuring} who presented the plus-minus statistic as a
regression problem. Doing so means `adjustments' can be made to the basic
plus-minus statistic to account for the strengths of the other players on a
team, and of the opposition players. The set up is again simple. Define a
segment of play to be one where the same set of players (usually two sides of 11
players) are on the pitch. A new segment is defined every time a new set of
players are on the pitch. This may occur when a substitution is made, or when a
sending off occurs, or for a different match. Each segment $t = 1,...,T$ is an
observation. The dependent variable is the goal difference
$\mathbf{y} = (y_1,...,y_T)$ per 90 minutes during segment $t$. Let there be $N$
players in total (in the whole league), then the independent variables form a
$T\times N$ design matrix $X$ of dummy variables defined as
 \[
  x_{tj}=\begin{cases}
               1 \text{ if player $j$ plays for the home team in the segment} \\
               -1 \text{ if player $j$ plays for the away team in the segment}\\
               0 \text{ if he doesn't play in the segment}
            \end{cases}
\]
where each player in the league is identified by a unique numeric index $j$.
The adjusted plus-minus statistic is
 then the solution to the regression model
 $y_t \sim \alpha X_t$, where $\alpha$ is an $n \times 1$ vector
 of parameters measuring the contribution of each player to the response variable (in
 this case, goal difference).

 In basketball, the number of segments within a game is much higher than the
 number players used in the game, and the matrix
 $\mathbf{X}^\intercal\mathbf{X}$ is `well-behaved' so that $\alpha$ can be
 estimated. In soccer however, the number of substitutions is limited to three
 per team (and are often not even used) and the number of segments is much
 smaller than the number of players on the pitch. Further, over the course of a
 match and season, the same groupings of players will play together. For
 example, a partnership between two centre backs is commonplace in soccer
 meaning they are on the pitch together for nearly every minute of play during
 an entire season. The result of all of this is that although the matrix
 $\mathbf{X}^\intercal\mathbf{X}$ is well-behaved for basketball, it is not so
 for soccer, and is singular, or near-singular, so that attempts to estimate
 $\alpha$ using ordinary least squares for example will fail.


 Ice-hockey suffers from these same problems and \citet{sill2010improved}
 presented a solution using ridge regularisation (also known as Tikhonov
 regularisation) instead of ordinary least squares to estimate the
 coefficients. The resulting methodology is known as the \emph{regularized
   adjusted plus-minus statistic}. Ridge regularisation is known to work well in
 the presence of collinearity and solves the problem by making a trade-off
 between minimising the estimation error (suppressing noise) and minimising the
 magnitude of the estimate (risking loss of information). In other words,
 instead of minimising the objective function in the usual squared errors
 problem:
\begin{align*}
 & \min ||\boldsymbol{\alpha} \boldsymbol{x} - \boldsymbol{y} ||^2_2 \\
  \boldsymbol{\alpha} & = (\boldsymbol{X}^\intercal\boldsymbol{X})^{-1}
                   \boldsymbol{X}^\intercal\boldsymbol{y},
\end{align*}
an alternative objective function, given by:
\begin{align*}
  & \min ||\boldsymbol{\alpha} \boldsymbol{x} - \boldsymbol{y} ||^2_2 +
    \lambda ||   \boldsymbol{x}||^2_2\\
  \boldsymbol{\alpha} & = (\boldsymbol{X}^\intercal\boldsymbol{X} +
                   \lambda^2 \boldsymbol{I})^{-1}
                   \boldsymbol{X}^\intercal\boldsymbol{y},
\end{align*}
is used. The penalty term, $\lambda$, penalizes large values of the parameters of interest.
The advantage of the ridge regression compared to other regularisation techniques such as the 
lasso for example is that it shrinks the coefficients of correlated predictors towards
each other whereas the lasso will tend to pick one and ignore the others. In the extreme
case of $k$ identical predictors, the ridge regularisation will give each of them
identical coefficients with $1/k$th the magnitude that any single one would get if were
the only one used as a covariate. This is very desirable in the case of estimating plus-minus 
ratings: if two players are always playing together (a pair of centre backs for
example), it is intuitively correct to say that their contributions to the team are
identical and thus award them identical ratings.

\section{New Plus-Minus Ratings for Soccer}
\label{sec:PlusMinusForFootball}
As a consequence of ice-hockey being a low scoring game, the latest developments
in the plus-minus metric have looked at using alternative dependent variables to
measure the team's success. The dependent variable is often called the `target'
as it is in some sense what the players should be targeting to improve during
the match. In ice-hockey, \citet*{macdonald2012expected} uses expected goals
rather than actual goals as the target variable, whilst
\citet*{macdonald2012adjusted} presents plus-minus models for shots. In this
section we present two new versions of the plus-minus metric: (a) a plus-minus
metric with difference in expected goals between the two teams as the target
variable, and (b) a plus-minus metric with change in expected points as the
target.

\subsection{Expected Goals Plus-Minus}
\label{sec:PMxG}

In recent years the concept of expected goals in soccer and ice-hockey has
become popular in the media (see, for example, \citet*{green2012}), In the
academic literature there has been limited interest with only exception being,
to the best of your knowledge, \citet*{lucey2014quality} and
\citet*{eggels2016expected}.

The idea behind the notion of expected goals (xG) is simple: for each shot on
goal that a team has, the expected number of goals is the probability of the
shot resulting in a goal. The probability of the shot being successful depends
on several factors: the location of the shot (proximity to the goal), the
player, the position of the defenders, the weather conditions, the fatigue of
the player, and so on. The reason xG has become a popular concept in soccer is
that it has been shown to be more informative than actual goals when judging how
well a team has played. Since goals are a rare event, they don't always reflect
properly a team's performance on the pitch. An alternative is to use shots,
which are an order of magnitude more common, instead of goals, but this has the
problem of considering all shots with equal standing, regardless of how good a
chance they have of being successful. An expected goals model deals with this
issue by assigning to each shot a measure of its quality, computed as the
probability the shot had of resulting in a goal.

In order to create our expected goals model, we compare the out-of-sample
performance of several probabilisitic classifiers trained on a large amount of
shots. Some of the earlier works have focused on finding expected goals models
that are as close as possible to the actual number of goals scored, which in our
opinion defeats the purpose of having a different more sophisticated
statistic. Instead, since we are interested in predicting an accurate
\textit{probability} that a given shot will result in a goal, we use
\textbf{Brier score loss} as the target for model training, hyperparameter
tuning, and cross-validation. A study in the same spirit was undertaken in
\citet*{lucey2014quality}, albeit they use \textbf{mean absolute error} as their
target metric.


Shots in football come from many different situations. We have separated our
shots into four different categories: \textbf{Penalty}, \textbf{Freekick},
\textbf{Header}, and \textbf{Open play}. The latter category contains all shots
taken with the foot that did not result directly from a set piece. Since the nature
of each of these types of shots is different, we designed our expected goals model
by fitting four \textit{specialist models}: one to each shot category. This means the
 feature selection process can be refined for each type of shot,
 and any redundant information is removed from the model 
  (for instance, there is no point on using shot
location when designing a model for penalties).

Our dataset contains over 600,000 shots event. Of those shots, roughly 61,000
resulted in a goal (a conversion rate of 10.2\%). The breakdown of shots by
types is in Table~\ref{table:shots} below. The baseline error is determined by
calculating the Brier score of a model predicting a constant probability (for
every shot type) equal to the empirical frequency of scoring a goal for that
particular shot type.

\begin{table}[ht]
  \centering
  \caption{Shot types and baseline errors (based on using the empirical frequency for that type of shot).}
  \begin{tabular}{lrrr}
    \toprule
    {}        &   Shots &  Goals &  Baseline error \\
    \midrule
    Free Kick     &   21,368 &   1,282 &        0.056 \\
    Header        &   99,620 &  11,438 &        0.102 \\
    Open Play     &  476,123 &  43,834 &        0.084 \\
    Penalty       &    6,498 &   4,912 &        0.185 \\
    \hline
    \textbf{Total}  &  603,609 &  61,466 &      0.091 \\
    \bottomrule
  \end{tabular}
\label{table:shots}
\end{table}

We consider the following features in order to train our models, all of them
normalized so that they have $[0,1]$ range.

\begin{itemize}
\item \textit{Horizontal pitch coordinate}: $x$, 1 corresponds to the goal line
  on the attacking side.
\item \textit{Adjusted vertical coordinate}: $y_{\text{adj}}$, 0 corresponds to
  a central position, 1 to either side of the pitch.
\item \textit{Goal view angle}: measuring the angle between the shot location
  and the two goalposts.
\item \textit{Inverse distance to goal}: measured to the
  center of the goal, 1 corresponds to the center of the goal, 0 to the furthest
  position on the pitch.
\item \textit{Time of play}: 0 being the kickoff and 1 corresponding to
  minute 90.
\item \textit{Goal value}: a measure of how the winning probability would be
  affected if a goal was scored, given as empirical frequencies, based on goal
  difference and game time remaining.
\item \textit{Big chance}: a boolean subjective indicator defined by Opta
  whenever a shot is deemed to be a very good chance, e.g. a one on one
  opportunity after a counterattack.
\item \textit{Goalkeeping skills}: for the opposition goalkeeper as detailed in
  Section~\ref{sec:data}.
\end{itemize}

It is worth noting that although there are EA SPORTS ratings for players' abillities to 
score goals, we do not include these in our expected goals models. This
is because the main purpose of our
expected goals model is to be used as a target for a plus-minus player rating,
and including information on the shooting player's ability would induce a feedback loop. 
One should also mention that although some of the features we consider are 
obviously  correlated (namely the pitch coordinates, the inverse distance to goal and 
the goal view angle) this relation is nonlinear, and therefore some families 
of classifiers benefit from the additional information.

We test four main families of machine learning models, \textbf{Logistic
  Regression}, \textbf{Random Forest}, \textbf{Gradient Boosting}, and
\textbf{Neural Network} (Multi-Layer Perceptron). In order to fine tune the
models' hyper-parameters, an inner-loop cross validation is performed on the
training set; the resulting model is then evaluated on the validation set in
order to get the out-of-sample score. Results are summarized in Table
\ref{table:xGmodels}.

\begin{table}[ht]
  \centering
  \caption{Summary of model errors for each shot type. The best performing model is highlighted in bold.}
  \begin{tabular}{lrrrr}
    \toprule
    {}            &   Penalty       & Free Kick     &  Header         & Open Play \\
    \midrule
    Baseline      &   0.1845      & 0.0564      &  0.1016       &  0.0836 \\
    Logistic Regression &   0.1847      & 0.0560        &  0.0927       &  0.0718 \\
    Random Forest   & 0.1845      & \textbf{0.0555} &  \textbf{0.0893}  &  0.0714 \\
    Gradient Boosting &   \textbf{0.1844} & 0.0556      &  0.0894       &  0.0714 \\
    Neural Network    &   0.1845      & 0.0564      &  0.0950       &  \textbf{0.0673} \\
    \bottomrule
  \end{tabular}
\label{table:xGmodels}
\end{table}

It is noticeable that there is no pattern - no one type of 
model always performs `best'. 
As a point of comparison, the mean absolute value from our combined best models 
is very similar to the one of the best model in \citet*{lucey2014quality}.
 However, the model used by \citet*{lucey2014quality} included 
 information on the position of the defending players on the opposition side.

It is worth noting some characteristics of the models for each shot type. 
Penalties require consideration separately to other shots. All penalties 
are taken from the same spot so shot location variables cannot be included in the model.
Further, only a few models manage to outperform the baseline score, and
the improvement is so small that is probably not significant. The bottom-line
here seems to be that the outcome of penalties are truly random, and therefore they should all 
be awarded the same value for expected goals, regardless of other considerations.

For the free-kick model, we find that the goalkeeper skill variables do not
seem to add any value to any of the models, with most of the predictive power
coming from the location based features. Similarly to penalties, the scores for
all the models trained are very close to the baseline.

The outcome of headed shots is heavily influenced by shot location, with the
goal view angle being the dominant variable in the model. Goalkeeper skill features seem to
have a minimal impact on the model performance and can be dropped without any
significant loss of performance.

By far the largest subset of shots in our dataset is open play shots. All the features seem
to add value to the models, with the exception of game time. The dominant features
are inverse goal distance, goal view angle, and the \textit{big chance}
indicator.

The resulting net expected goals for each segment of play (in which the same set of 22 players is
on the pitch) is used as the dependent variable (or target) in out expected goals plus-minus (xGPM) player rating.

\subsection{Expected Points Plus-Minus}
\label{sec:PMxP}
The ultimate objective of a soccer match is to win.
Team managers and fans want to know which players perform well when
the match is tight and the scoreline is close. Using the regularized
adjusted plus-minus metric, or the xG plus-minus metric presented
above, does not account for the match situation. As such, we propose a
new plus-minus metric based on expected points. In soccer leagues
around the world, a team is awarded 3 points for a win, 1 point for a
draw and 0 points for a loss. The expected points for the home team in minute $t$ of a match is then
\[
xP^H_t = 3 \times \Prob^{HW}_t + 1\times \Prob^{D}_t,
\]
where $\Prob^{HW}_t$ is the probability of the home team winning the match
evaluated at time $t$, taking into account the current scoreline and
the number of players on each team. $\Prob^{D}_t$ is the probability
of the team drawing the match evaluated at time $t$.

In calculating our new plus-minus expected points statistic, 
we compare the expected points at the start of a segment
of play with those at the end of a segment of play. For example,
suppose that the first change in team lineups in a particular match happened in minute 60
(through substitution(s) or a red card dismissal(s)).
 The change in expected points for the home team is
$\Delta xP^H = xP^H_{60} - xP^H_{0}$, whilst the change in expected points for the away team is
$\Delta xP^A = xP^A_{60} - xP^A_0$. 
The target variable we propose is then the change in expected points 
for the home team minus the change in expected points for the away team,
$y_{[0, 60]} = \Delta xP^H - \Delta xP^A$.

In order to calculate expected points variables we need an `in-play' model
to estimate the probabilities of the home team winning, a draw 
and the away team winning at any moment of the match.
The model used here is a simplification of the random point process model described in
\citet*{volf2009random}. This process is fully
characterised by the scoring intensity functions (also known as hazards)
of the home and away teams
 $\lambda_H(t)$, and $\lambda_A(t), t \geq 0$ which are non-negative, bounded,
measurable functions of $t$. The
intensity is allowed to depend on some covariates $Z(t)$.
$Z(t)$ is in turn an observed random process that can depend on
time. A common framework to model the effect of covariates on the
intensity function is to use a proportional hazard model, first
described in \citep*{Cox1962}.

Here, the hazards of each team scoring depend on 
two categorical covariates describing the game context at 
time $t$. They are defined by:
\begin{itemize}
\item $z_{GD}(t) = {-3 \leq, -2, \dots, 2, \geq 3}$ defines the goal
  differential with respect to the home team. We found that a
  truncation at $3$ goals difference works well in practice.
  
\item $z_{MP}(t) =  {-3 \leq, -2, \dots, 2, \geq 3}$ defines the man
  power advantage with respect to the home team. 
\end{itemize}

The model basically assumes that each team scores goals at a rate that depends
on the time of the match, the number of red cards received by the two teams, and
home advantage.  The simplification we adopt over \citet*{volf2009random} is to
not take account of the strengths of the two teams playing in any particular
match. As such, we are effectively using `average' probabilities over all
games. The justification for this is again very similar to what we argue in
Section \ref{sec:PMxG}; the identities of the players are already being taken
into account in the model and accounting for them again in the calculation of the
in-play probabilities is in some sense double counting, and results in
`punishing' players on good teams with high probabilities of winning matches.

The initial (average) probabilities of a home win, a draw and an away win at
$t=0$ can be computed from the empirical frequency. Using the last eight years
of results from the English Premier League, these probabilities are 0.46, 0.26
and 0.28 respectively. The corresponding expected points at $t=0$ are then 1.63
for the home team and 1.11 for the away team. We computed similar quantities for
every league we have in our data.

Returning to our example, we can calculate the target variable as
\begin{align}
  y_{[0, 60]} &= (xP^H_{60} - xP^H_{0}) - (xP^A_{60} - xP^A_{0}) \\
  & = (3 \times \Prob^{HW}_{60} + \Prob^D_{60} - 3 \times \Prob^{HW}_{0} -
    \Prob^D_{0}) - (3 \times \Prob^{AW}_{60} + \Prob^D_{60} - 3 \times \Prob^{AW}_{0} -
    \Prob^D_{0})  
\end{align}  
The model computes these probabilities and the
 corresponding target $y_{[0, 60]}$ for this game segment can be computed.
This model is fitted as explained in \citet*[Section
4]{volf2009random}, and estimated probabilities are obtained by
simulation using the procedure detailed in \citet*[Section 5]{volf2009random}.

This new target directly rewards players for
contributing to the final result. Previous plus-minus ratings, including the expected goals plus-minus rating
described above credits players for creating chances and scoring goals irrespective of the 
influence of them on the final result.

\subsection{Minor Modifications to Plus-Minus Ratings for Soccer}
\label{ssec:minorModif}
\subsubsection*{Adjusting for Man Power}
The effect of receiving a red card has been studied in soccer (see, for example,
\citet*{ridder1994down} and \citet*{liu2016modelling}) and has been found to be
beneficial for the opposing team in terms of scoring rate. Further, the
advantage is larger in the case of the home team benefiting from having more
players on the pitch.

In ice-hockey, the effect of player expulsion in plus-minus ratings has been
modelled using a situation specific coefficient for each player: a coefficient
for even-strength situations and another one during shorthanded situations
\citep*{macdonald2011improved}.  This solution has the effect of doubling the
number of estimated coefficients and is not suitable for large numbers of
players, and given the extremely low frequency of red cards in soccer, is
unnecessary.

The solution we propose here is different. We describe the effect of receiving a red card
using a dummy variable capturing the average penalty suffered by a team with one (or more)
man down. When a team is shown its first red card, the player in question is replaced by
the `first dismissal' dummy player. A second dismissal leads to the substitution of the
offending player for a `second dismissal' dummy, and so forth. However, for each dismissal
that is `cancelled out' when a team loses one of its `surplus' players, the relevant
dismissal dummy is reset to 0. We use three \emph{dismissal} dummy variables to cover the
maximum number of dismissals occurring in the data.


\subsubsection*{Home Advantage}
Home advantage in soccer was first discussed in the academic literature by
\citet*{pollard1986home} and many researchers have since measured its magnitude
\citep*{clarke1995home} and tried to explain its variation over time
\citep*{pollard2005long} and space \citep*{pollard2006worldwide}.

When computing the plus-minus statistic for basketball
\citet*{winston2012mathletics} accounted for home advantage by adjusting the
points differential (the dependent variable in the regression model) by the
average number of points by which the home team defeats the away team (3.2 per
48 minutes). Rather than adjusting the dependent variable, the solution we
propose here is to add an intercept term to the regression problem which
represents the average home advantage over all teams in the competition. This is
more in line with what has been done previously in the soccer literature (see,
for example, \citet*{maher1982modelling}, \citet*{dixon1997modelling},
\citet*{koopman2015dynamic}, \citet*{boshnakov2017bivariate}).


\subsubsection*{Chronology of Performances}
It is widely accepted in sports that recent performances are more informative
when predicting future performances.  Therefore, in order to increase the
predictive power of our rating, we apply a weighting scheme to the different
observations (segments) when fitting the ridge regression.  The weights are
computed as follows:
\[
    w_i = \text{exp} \Big ( \zeta (\text{date}_i - \text{ratingDate}) / 3.5 \Big )
\]
with $\zeta$ being the time-weighting parameter ($\zeta = 0$ corresponds to
the non-weighted regression), $\text{date}_i$ the date of the $i$th
observation (segment) and $\text{ratingDate}$ is the date when the
rating is computed. Following standard practise in soccer modelling
(\citep*{dixon1997modelling}; \citep*{boshnakov2017bivariate}), time distances
are scaled in half week units.

\subsubsection*{League Competitiveness}
Since we have data covering several leagues across Europe, we must control for
any differences in strengths of the leagues themselves. For example, some
leagues may have stronger players on average than other leagues. Two players of
equal ability will perform differently if one is in a strong league whilst the
other players in a weak league. The Union of European Football Associations
(UEFA) itself acknowledges the inequity of ability across leagues and publishes
a ranking by country and awards slots in European competitions accordingly. The
consequence on our ratings of this is that a bias could be introduced so that
players in weak leagues have inflated ratings. This problem appears when data
from various competitions are used to fit the Ridge regression.

We correct for this bias by using the players
traveling between leagues to compare the strengths of each league. To do so,
we introduce one coefficient $x_l$ per league in the data. Assume we have $L$
leagues and let $m_{il}$ be the number of home team players
minus the number of away team players, considering only players whose
considered at time of match $i$ to be adapted to competition $l, l = 1, \dots, L$.
A player is considered adapted to a competition if he plays at least $6$ games
in the current season in that competition or if he played more games in
this competition than in any other over the previous 18 months to the game date.
Hence, $x_l$ is the weight of $m_{il}$ in the Ridge regression and represents the
adjustment we need to apply to a player joining a new league.

The final Ridge regression will need to estimate $N + 1 + 3 + L$ parameters ($N$
players, a home advantage parameter, three dismissal parameters, 
and $L$ league parameters). The
model's design matrix is very sparse with a limited number of non zeros
entries per row. The model also has two hyper-parameters (the Ridge penalty
$\lambda $ and the time weighting $\zeta $) which need to be fine-tuned using
cross-validation.

\section{Computation Results and Discussion}
\label{sec:Testing}
\subsection{Computation}
The game segmentation algorithm (Section~\ref{ssec:RegAdjustedPM}) as well as
the minor adjustments described in Section~\ref{ssec:minorModif} are applied to
the data described in Table~\ref{tab:EuroLeague} using \citet{Rlang} and the
result is stored in sparse matrix object implemented in the contributed package
\texttt{Matrix} \citep{MatrixPkg}. The computation resulted in $129,988$
segments and $N = 10,983$ players' ratings to be estimated. The Ridge regression
was performed using the contributed package \texttt{glmnet} \citep{glmnetPkg}
and a multi-response Gaussian model using a ``group'' penalty on the
coefficients for each variable (also known as multi-task learning).

\subsection{Results}
\subsubsection{Model Tuning}
As mentioned in Section~\ref{sec:PlusMinusForFootball}, the model has two hyper-parameters
namely the Ridge penalty $\lambda$ and the time weighting $\zeta$ which need to
be fine-tuned. The strategy adopted here is to use the new PM player ratings in
an ordered probit regression model to predict the match outcomes (home win/draw/
away win) and use the value of the hyper-parameters that minimised the
out-sample Brier score. A 10-fold cross-validation was used to split the data
into training and testing sets and the process was repeated three times. The
covariates used are the average PM ratings derived in
Section~\ref{sec:PlusMinusForFootball} for the starting 11 players using data
from the two years prior to the game date\footnote{Different length windows were
  tried and two years was found to perform best in terms of Brier score.}.  The
best model achieved an average Brier score of $0.292$ (sd = $0.003$) which is
similar to the accuracy achieved by the market for the same set of
games\footnote{The adjusted probabilities deduced from bet365 pre-match betting
  odds achieved a Brier score of $0.295$, after removing the bookmaker vigorish,
  for the same set of games.} with $\mathbf{\lambda = 0.042}$ and
$\mathbf{\zeta = 0.002}$.  Note that the defintion of Brier score apopted here
follows the original formulation given by \citet{brier1950verification} and is
defined by $BS = \frac{1}{N} \sum_{i=1}^N \sum_{i=1}^R {(p_{ti} - o_{ti})}^2$ in
which $p_{ti}$ is the probability that was forecast for outcome $i$, $o_{ti}$ is
the dummy variable equal to one if outcome $i$ is observed and $R = 3$.

Expected goals models from Section~\ref{sec:PMxG} were fitted using 10-fold
cross-validation, with hyper-parameter tuning in the inner loop. Logistic
regression and random forest models used the implementation in \texttt{scikit-
  learn} \citep{scikit-learn}. Gradient boosing models were fitted and tuned
using \texttt{xgboost} \citep{xgboost}. Neural network models were fitted and
tuned using \texttt{Keras} \citep{keras}. Plots in Figure~\ref{fig1} were 
generated using \texttt{matplotlib} \citep{matplotlib}. 

\subsubsection{Fitting Results}
Before we investigate the actual players ratings, we study in this section the
significance of the other adjustments we introduced in
Section~\ref{ssec:minorModif}, namely \emph{man-power} and \emph{home
  advantage}. The Ridge regression was fitted using the last two seasons and the
results are summarised in Table~\ref{tab:fitRes}.
\begin{table}[ht]
  \centering
  \caption{Impact of red cards on the three plus-minus ratings.}
  \begin{tabular}{lrrr}
    \toprule
    Parameter     &  PM &  xGPM &  xPPM \\
    \midrule
    Red Card 1    &  -1.25   & -1.18  &  -0.12 \\
    Red Card 2      &  -0.16   & -0.15  &  -0.01 \\
    Red Card 3    &  -0.012  & -0.005 &  -0.001 \\
    \hline
    home Advantage      & 0.006    & 0.005  & 0.0004 \\
    \bottomrule
  \end{tabular}
\label{tab:fitRes}
\end{table}
The first red card has a large negative effect on all three ratings, whereas
additional dismissals contribute a much smaller effect. One explanation is that
a first red card is very likely to be followed by a considerable change in team
tactics, and may happen early enough in a match to leave the opposing team with
enough time to take advantage of the extra man-power.  Further reductions will
have an added negative effect, but will not be associated with a further change
in tactics, and are very likely to occur late on in a game, when there is less
time to change the match result.

The estimated home advantage effect is surprisingly very small for the goal
based and expected shot based PM rating and almost zero for the xPPM one
suggesting that players do not perform, on average, differently playing home or
away. It is worth noting here that finding a home advantage of zero for the xPPM
rating is expected as we have already accounted for it when setting the initial
expected points as explained in Section~\ref{sec:PMxP}.

\subsubsection{Player ratings evolution}

\begin{figure}[ht] 
  \begin{subfigure}[b]{0.55\linewidth}
    \centering
    \includegraphics[width=1\linewidth]{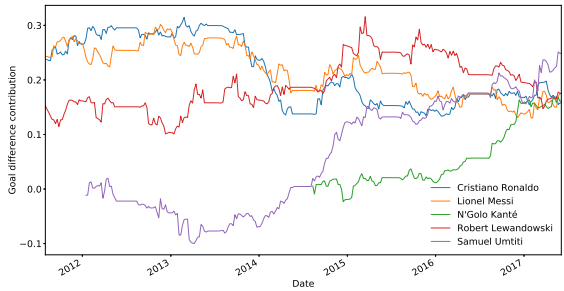} 
    \caption{Actual goals plus-minus.} 
    \label{fig1:a} 
    \vspace{4ex}
  \end{subfigure}
  \begin{subfigure}[b]{0.55\linewidth}
    \centering
    \includegraphics[width=1\linewidth]{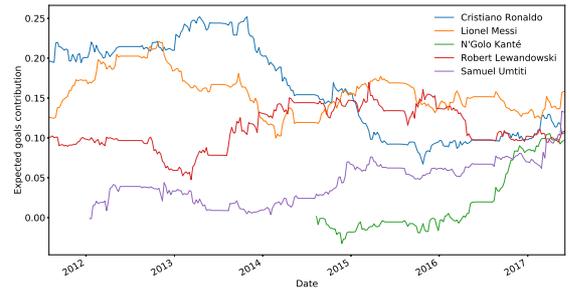} 
    \caption{Expected goals plus-minus, xGPM.} 
    \label{fig1:b} 
    \vspace{4ex}
  \end{subfigure} 
  \begin{subfigure}[b]{0.55\linewidth}
    \centering
    \includegraphics[width=1\linewidth]{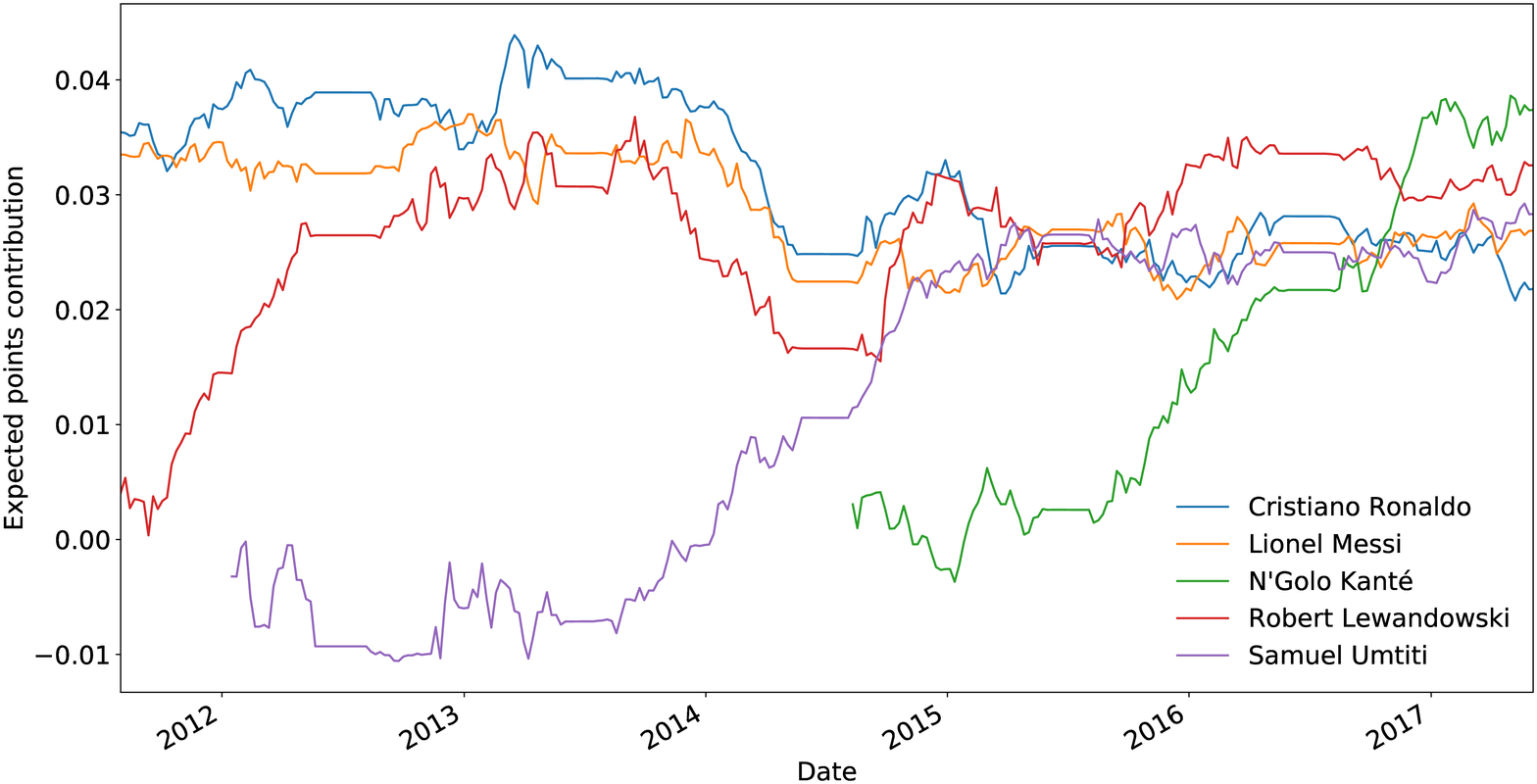} 
    \caption{Expected points plus-minus, xPPM.} 
    \label{fig1:c} 
  \end{subfigure}
  \begin{subfigure}[b]{0.55\linewidth}
    \centering
    \includegraphics[width=1\linewidth]{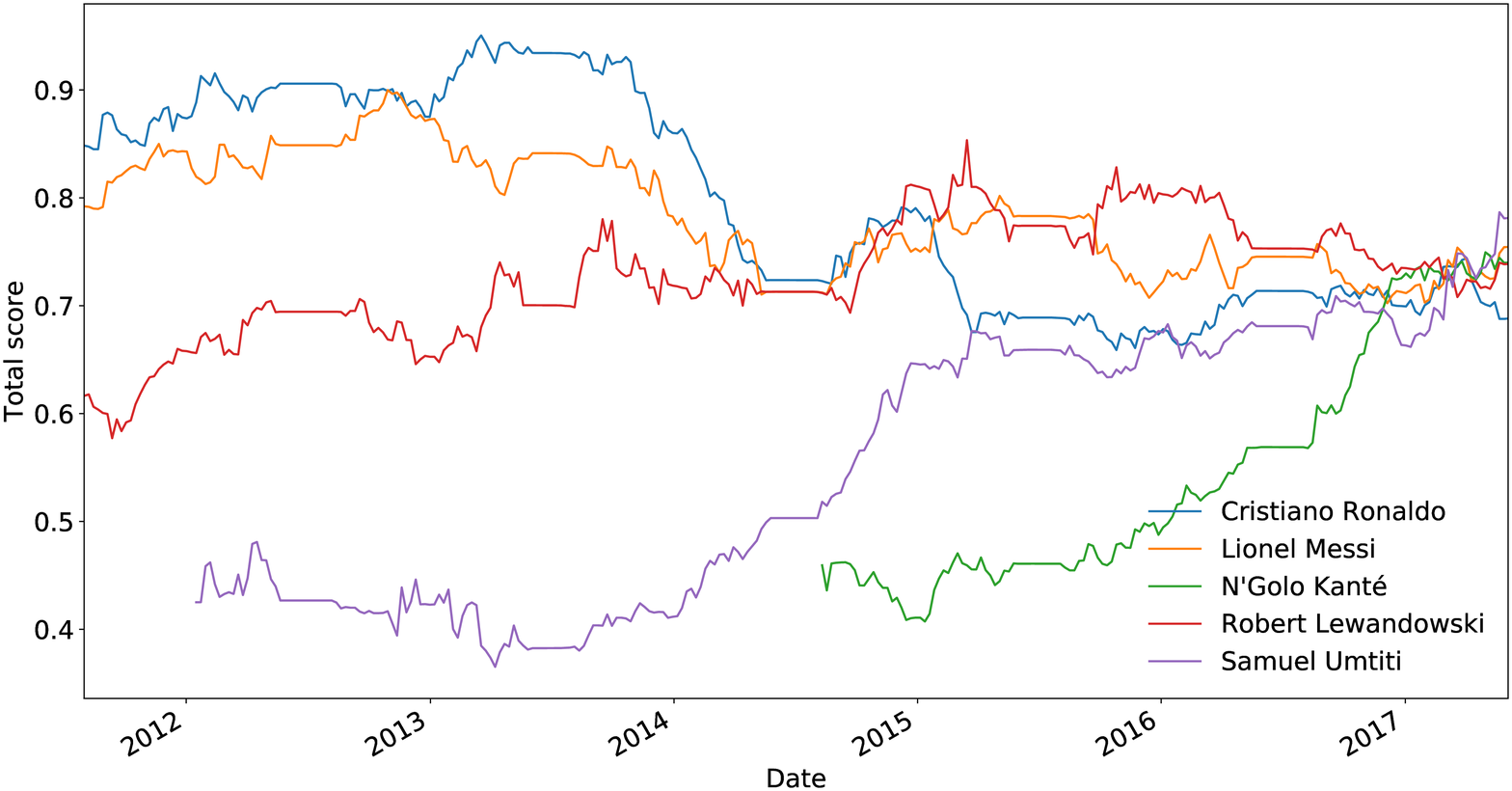} 
    \caption{Average of the three plus-minus ratings.} 
    \label{fig1:d} 
  \end{subfigure} 
  \caption{Evolution of the three plus-minus ratings for five example players.}
  \label{fig1} 
\end{figure}

%
One of the interesting applications of the plus-minus ratings is observing how
the performance of selected players has evolved over time. In Figure~\ref{fig1}
we plot the relative contributions of Cristiano Ronaldo, Lionel Messi, N'Golo
Kant\'e, Robert Lewandowski and Samuel Umtiti to their respective teams
according to each of the three plus-minus ratings: goals PM, xGPM, and xPPM.

We can observe how the three forwards experienced a decline during the 2013/14
season (perhaps they were conserving energy ahead of the 2014 World Cup?), and
how the stellar performance of N'Golo Kant\'e over the last two seasons pushed
him to the top of the ratings. In fact, Kant\'e is top according to both the
goals PM and the xPPM, whereas Lionel Messi tops the ratings for xGPM. We also
show the mean of the three normalized ratings for illustration and the five
players are very similar in value.

\subsubsection{Challenging the Ballon D'or Results}
The Ballon d'Or\footnote{https://en.wikipedia.org/wiki/Ballon\_d\%27Or} is the
most prestigious individual distinction in soccer and is awarded to the player
deemed to have performed the best over the previous calendar year, based on
voting by expert journalists.

\begin{table}[ht]
  \centering
  \begin{tabular}{lllr}
  \toprule
  Year & Position &            Player &  Score \\
  \midrule
  2011 & 1 &         Andr\'es Iniesta &  0.915 \\
       & 2 &        Cristiano Ronaldo &  0.914 \\
       & 3 &                    Pedro &  0.906 \\
  \midrule
  2012 & 1 &              Eden Hazard &  0.884 \\
       & 2 &            Mirko Vucinic &  0.873 \\
       & 3 &             Lionel Messi &  0.847 \\
  \midrule
  2013 & 1 &           Gerard Piqu\'e &  0.935 \\
       & 2 &        Cristiano Ronaldo &  0.904 \\
       & 3 &      Francesc F\'abregas &  0.901 \\
  \midrule
  2014 & 1 &             Manuel Neuer &  0.918 \\
       & 2 &           Jerome Boateng &  0.882 \\
       & 3 &             Lionel Messi &  0.869 \\
  \midrule
  2015 & 1 &              David Alaba &  0.944 \\
       & 2 &       Robert Lewandowski &  0.897 \\
       & 3 &           Edinson Cavani &  0.888 \\
  \midrule
  2016 & 1 &           N'Golo Kant\'e &  0.915 \\
       & 2 &            Claudio Bravo &  0.896 \\
       & 3 &            Luis Su\'arez &  0.890 \\
  \bottomrule
  \end{tabular}
  \caption{Top player scores per calendar year}\label{table:topPlayers}
\end{table}

The plus-minus ratings provide us with an alternative way to make a top-player
classification for every calendar year. As a proof of concept, we have computed
the average of the three variations of the plus-minus rating, each of them
previously normalized to the $[0, 1]$ range, and filtered out players who didn't
play at least 900 minutes (the equivalent of 10 full games). The results are
summarized in Table~\ref{table:topPlayers}. Despite the fact that the Ballon
d'Or award has been dominated by Lionel Messi and Cristiano Ronaldo over the
last years\footnote{In fact, one needs to go as far back as 2007 to find a
  Ballon d'Or that was \textbf{not} awarded to either Messi or Ronaldo!} our
scores suggest that perhaps some other players might have deserved the
recognition.

\subsubsection{Comparing League Strengths}
In this section we examine the results of adjusting for league strength in the
PM ratings (Section~\ref{ssec:minorModif}).  The parameter estimates can be
used to compare league strength and this can be used to help clubs understand
how players from other leagues might perform in their league. The results are
again normalized to the $[0, 1]$ range and summarized for each of our PM ratings
in Table~\ref{tab:leagueComp}. The final column of the table shows the mean
league strength parameter estimate.
 
\begin{table}[ht]
\centering
\begin{tabular}{rlrrrr}
  \toprule
 & competitionName & PM & xGPM & xPPM & meanPM \\ 
  \midrule
1 & England Premier League & 1.00 & 0.67 & 0.97 & 0.88 \\ 
  2 & Germany Bundesliga & 0.92 & 0.32 & 1.00 & 0.75 \\ 
  3 & Spain La Liga & 0.43 & 1.00 & 0.49 & 0.64 \\ 
  4 & Italy Serie A & 0.61 & 0.64 & 0.66 & 0.64 \\ 
  5 & Russia Premier League & 0.49 & 0.52 & 0.61 & 0.54 \\ 
  6 & Germany Bundesliga II & 0.55 & 0.18 & 0.86 & 0.53 \\ 
  7 & England Championship & 0.63 & 0.27 & 0.53 & 0.48 \\ 
  8 & Portugal Liga NOS & 0.69 & 0.00 & 0.49 & 0.39 \\ 
  9 & France Ligue 1 & 0.25 & 0.45 & 0.18 & 0.29 \\ 
  10 & Turkey Super Lig & 0.12 & 0.10 & 0.38 & 0.20 \\ 
  11 & Netherlands Eredivisie & 0.00 & 0.32 & 0.00 & 0.11 \\ 
   \bottomrule
\end{tabular}
\caption{League Ranking according to the PM rankings.}
\label{tab:leagueComp}
\end{table}

The English Premier League dominates the ranking with high scores in goals and
points based PM ratings. The second strongest league appears to be the German
Bundesliga. The Spanish league scores the highest in terms of expected goals PM
but is slightly behind in terms of goals and expected points which may suggest
that players trained in this league have a worse conversion ratio (converting
opportunities to goals).  Surprisingly, the second divisions in Germany and
England seem to perform better than the top division in France and Portugal. One
possible explanation for this result is that teams get promoted from the second
tier divisions in Germany and England and perform better in the top leagues than
the players moving from Ligue 1 into these leagues. This may be a result of the
players from the second tiers divisions being more familiar with the environment
as they have not had to move countries to move leagues.  Netherlands seem to be
the `weakest' league among the set of leagues we analysed.

\section{Conclusions}
\label{sec:conclusion}
The paper presents a plus-minus ratings system adapted to soccer. We have
proposed two new versions of the plus-minus model designed to react to
particular aspects of the game.  Our first new plus-minus rating identifies
players who change the net expected goals of a team. We have called this the
expected goals plus-minus rating, xGPM. The second new plus-minus rating we
propose is designed to identify players who change the results of teams by
affecting the expected points of a team. We call this rating the expected points
plus-minus rating, xPPM.

We have used the new ratings to identify potential alternatives to the Ballon
d'Or winner - an award given each year to the best footballer on the planet. We
have also used the ratings to examine the evolution of five players'
performances over time. The rise of N'Golo Kante is quite remarkable, and during
the 2016-17 season, the models suggest he was the top player in our data (which
covers the top leagues in world football). Lastly, we used the model to estimate
the relative strengths of the leagues. It appears the English Premier League is
slightly stronger than the German Bundesliga, followed by Spain's La Liga.
Somewhat surprisingly, France Ligue 1 is rated as weaker than the second tier
divisions in both England and Germany.

Future work may look at using these ratings as part of a forecasting model for
match results. Alternatively, to aid those who make decisions regarding team
lineups, one could investigate how pairings of players perform together. For
example, a coach may be interested in knowing which central defensive pairing is
the most effective. For now, we hope that the objectivity of these new ratings
and the seemingly `expected' results may mean that plus-minus ratings are used
more readily in the soccer industry - both by clubs, fans and the media.

\section*{Acknowledgements}
We would like to thank Rick Parry for pointing us in the direction of plus-minus ratings.

\end{document}